\documentclass[showpacs,aps,prb,10pt]{revtex4}

\begin{document}


\title{Diffuse transport and spin accumulation 
in a Rashba two-dimensional electron gas }

\author{Jun-ichiro Inoue}
\affiliation{Department of Applied Physics, Nagoya University, 
Nagoya 464-8603, Japan}
\affiliation{CREST, JST, Japan}

\author{Gerrit E.W.\ Bauer}
\affiliation{Department of Applied Physics and DIMES, 
Delft University of Technology, Lorentzweg 1, 2628 CJ Delft, The Netherlands}

\author{Laurens W.\ Molenkamp}
\affiliation{Physikalisches Institut (EP3), Universit\"at W\"urzburg, 
D-97074 W\"urzburg, Germany}

\date{\today}

\begin{abstract}
The Rashba Hamiltonian describes the splitting of the conduction band as a
result of spin-orbit coupling in the presence of an asymmetric confinement
potential and is commonly used to model the electronic structure of confined
narrow-gap semiconductors. Due to the mixing of spin states some care has to
be exercised in the calculation of transport properties. We derive the
diffusive conductance tensor for a disordered two-dimensional electron gas
with spin-orbit interaction and show that the applied bias induces a spin
accumulation, but that the electric current is not spin-polarized.
\end{abstract}


\maketitle

Utilizing the spin degree of freedom for electronic applications is a
declared goal of the research field of magnetoelectronics or spin-electronics%
\cite{wolf}. Devices made from\ metallic layered systems displaying the
giant \cite{baibich} and tunnel magnetoresistance \cite{miyazaki,moodera}
have been proven useful for read-head sensors and magnetic random access
memories. Integration of such devices with semiconductor electronics is
desirable but turned out to be difficult because a large resistivity
mismatch between magnetic and normal materials is detrimental to spin
injection\cite{schmidt1}. Still, this problem can been solved in various
ways and spin injection into bulk semiconductors has indeed been reported%
\cite{fiederling,ohno,zhu,nrl,deboeck,schmidt2}. Electrical spin injection
into a high-mobility two-dimensional electron gas (2DEG) and its detection
appears to be much more demanding\cite{hammar4}. In this context it would be
attractive if application of an electric field alone would suffice to induce
a non-equilibrium magnetization or spin accumulation in the presence of the
spin-orbit interaction. Such an effect, dubbed the ``kinetic magnetoelectric
effect''\ was actually predicted in seminal theoretical work by Levitov $%
et\;al.$\cite{levitov}, and is caused by the combined action of the
spin-orbit interaction, absence of inversion symmetry, and the time-reversal
symmetry-breaking by an electric field in disordered systems. In asymmetric
heterostructures made from narrow-gap semiconductors the spin-orbit
interaction is dominated by the so-called Rashba term\cite{rashba}, which
has a very simple structure, can be quite significant\cite{nitta}, and
modulated by gate fields \cite{datta}. Recent observations of a
spin-galvanic effect\cite{ganichev} and spin-orbit scattering induced
localization/antilocalization transition in 2DEGs\cite{Miller,koga} reflect
the interest and importance of the topic.

Edelstein\cite{edelstein}{\it \ }showed that an applied field induces an
in-plane magnetization in a Rashba 2DEG. Although he did not make any
suggestions in this direction, the interpretation of the spin accumulation
being caused by an effective magnetic field has lead subsequently to the
misconception that the current is also spin polarized. A microscopic
calculation of both spin accumulation and current (or conductivity tensor)
on an equal footing is thus required. Furthermore, the existence of the
spin-orbit induced spin-accumulation and the conditions for its
observability have recently been a matter of controversy\cite%
{hammar,hammar2,hammar3,monzon,wees}.

In this Report we carry out microscopic model calculations of the
conductivity tensor and spin accumulation for a disordered Rashba 2DEG in
the linear response regime. This task is complicated by the correction to
the electric field vertex, which does not vanish even for short-range
isotropic scatterers. We confirm that a spin-accumulation normal to the
applied electric field vector is excited. However, the electric current is
not spin polarized, thus solving the controversies mentioned above. We
furthermore show that the mobility increases quadratically with the Rashba
spin-orbit interaction.

The Rashba Hamiltonian in the momentum representation and Pauli spin space
reads 
\begin{equation}
H_{0}=\left( 
\begin{array}{cc}
\frac{\hbar ^{2}}{2m}k^{2} & i\langle \alpha E_{z}\rangle k_{-} \\ 
-i\langle \alpha E_{z}\rangle k_{+} & \frac{\hbar ^{2}}{2m}k^{2}%
\end{array}%
\right) ,
\end{equation}%
where $k_{\pm }=k_{x}\pm ik_{y}\ $with ${\bf k=}\left( k_{x},k_{y}\right) $
the electron momentum in the 2DEG plane. $\langle \alpha E_{z}\rangle \ $%
parametrizes the spin-orbit coupling and is experimentally accessible\cite%
{nitta}. The eigenfunctions of the Hamiltonian are 
\begin{equation}
\phi _{{\bf k}s}=\frac{1}{\sqrt{2L^{2}}}e^{i{\bf k}\cdot {\bf r}}%
{is\frac{k_{-}}{k} \choose 1}%
,
\end{equation}%
with $s=\pm $ , $k=\sqrt{k_{x}^{2}+k_{y}^{2}}$ and $L^{2}$ the area of the
2DEG and corresponding eigenvalues are given as $E_{ks}=\hbar ^{2}k^{2}/2m+s$
$\langle \alpha E_{z}\rangle k.$ The current operator in Pauli spin space is
given as 
\begin{equation}
j_{x}=ev_{x}=e\left( 
\begin{array}{cc}
bk_{x} & i\lambda \\ 
-i\lambda & bk_{x}%
\end{array}%
\right) ,
\end{equation}

\begin{equation}
j_{y}=ev_{y}=e\left( 
\begin{array}{cc}
bk_{y} & \lambda \\ 
\lambda & bk_{y}%
\end{array}%
\right) ,
\end{equation}%
with $b=\hbar /m$ and $\lambda =\langle \alpha E_{z}\rangle /\hbar$. In
the space of the eigenfunctions of $H_{0}$, referred to hereafter as the $s$%
-space ($s=\pm $), the current operators are transformed as 
\begin{equation}
J_{x(y)}=U^{\dagger }j_{x(y)}U = e \left( bk_{x(y)}{\bf 1} 
+ \lambda \frac{k_{x(y)}}{k}{\bf \sigma }_{z} 
-(+) \lambda \frac{k_{y(x)}}{k}{\bf \sigma }_{y} \right) \,,
\end{equation}%
by the unitary matrix 
\begin{equation}
U=\frac{1}{\sqrt{2}}\left( 
\begin{array}{cc}
i\frac{k_{-}}{k} & -i\frac{k_{-}}{k} \\ 
1 & 1%
\end{array}%
\right) .
\end{equation}%
We also need the transformed spin matrices, $U^{\dagger }\sigma _{x}U,$ $%
U^{\dagger }\sigma _{y}U$ and $U^{\dagger }\sigma _{z}U$ to evaluate the
spin accumulation.

The standard model for disorder consists of randomly distributed, identical
point defects, which are neither spin-dependent nor flip the spin: 
\begin{equation}
V({\bf r})=V{\bf 1}\sum_{i}\delta ({\bf r}-{\bf R}_{i}).
\end{equation}%
In the following, we expand the Green function $G=(z{\bf 1}-H)^{-1},$ where $%
H=H_{0}+V({\bf r})$ and $z=\epsilon \pm i\eta ,$ in terms of the unperturbed
Green function, $G_{0}=[z{\bf 1}-H_{0}]^{-1},$ with matrix elements $g_{k\pm
}=1/(z-E_{k\pm })$ in $s$-space, and calculate the self-energy $\Sigma $ in
the Born approximation.

After ensemble averaging over the impurity distribution, denoted by $\langle
\cdots \rangle _{AV},$ and disregarding a trivial constant term, the
self-energy in the Born approximation reads: 
\begin{eqnarray}
\left\langle \langle {\bf k}s|VG_{0}V|{\bf k}^{\prime \prime }s^{\prime
\prime }\rangle \right\rangle _{AV} &=&\frac{nV^{2}}{4L^{2}}\delta _{{\bf kk}%
^{\prime \prime }}\sum_{k^{\prime }s^{\prime }}g_{k^{\prime }s^{\prime
}}\left( 1+ss^{\prime \prime }+ss^{\prime }\frac{k_{+}k_{-}^{\prime }}{%
kk^{\prime }}+s^{\prime }s^{\prime \prime }\frac{k_{+}^{\prime }k_{-}}{%
k^{\prime }k}\right) \\
&=&\frac{nV^{2}}{2L^{2}}\delta _{{\bf kk}^{\prime \prime }}\delta
_{ss^{\prime \prime }}\sum_{k^{\prime }s^{\prime }}g_{k^{\prime }s^{\prime
}}=\Sigma \delta _{{\bf kk}^{\prime \prime }}\delta _{ss^{\prime \prime }},
\end{eqnarray}%
where $n\equiv N/L^{2}$ is the density of impurities per unit area. Eq. (9)
follows from the odd symmetry of $k_{+}k_{-\text{ \ }}^{\prime }$ and $%
k_{+}^{\prime }k_{-}$ with respect to $k_{x}^{\prime }$ or $k_{y}^{\prime }$%
. The Green function is therefore given as 
\begin{equation}
\left\langle \langle {\bf k}s|G|{\bf k}^{\prime \prime }s^{\prime \prime
}\rangle \right\rangle _{AV}=\frac{1}{g_{ks}^{-1}-\Sigma }\delta _{{\bf kk}%
^{\prime \prime }}\delta _{ss^{\prime \prime }}=\tilde{G}_{ks}.
\end{equation}%
We find both $\tilde{G}$ and $\Sigma $ to be diagonal in ${\bf k}$ and $s$.
By direct inspection it can be seen that $\langle VG_{0}VG_{0}VG_{0}V\rangle 
$ is diagonal, meaning that the {\em exact} self-energy must be diagonal as
well.

The (longitudinal) conductivity is given by the Kubo formula as 
\begin{equation}
\sigma _{xx}=\frac{\hbar }{4\pi L^{2}}{\rm Tr}\left\langle
J_{x}G^{A}J_{x}G^{R}+J_{x}G^{R}J_{x}G^{A}\right\rangle _{AV},
\end{equation}%
where the superscripts $A$ and $R$ denote ${\it Advanced}$ and ${\it Retarded%
}$, respectively, and will be omitted below for brevity. We evaluate $%
\langle J_{x}GJ_{x}G\rangle _{AV}=J_{x}\langle GJ_{x}G\rangle _{AV}\equiv
J_{x}K$ in the ladder approximation, which obeys the Ward relation with the
self-energy in the Born approximation: 
\begin{equation}
K\sim \tilde{G}J_{x}\tilde{G}+\tilde{G}\langle VKV\rangle _{AV}\tilde{G}.
\label{vertex}
\end{equation}%
The matrix elements of $\langle VKV\rangle _{AV}$\ \ are 
\begin{eqnarray}
\langle \langle ks|VKV|k^{\prime }s^{\prime }\rangle \rangle _{AV}
&=&\sum_{k_{1}s_{1}}\sum_{k_{2}s_{2}}\left( \frac{V}{2L^{2}}\right)
^{2}\sum_{i}\langle e^{-i({\bf k-k}_{1})\cdot {\bf R}_{i}}e^{i({\bf k}%
^{\prime }-{\bf k}_{2})\cdot {\bf R}_{i}}\rangle _{AV}  \nonumber \\
&\times &\left( 1+ss_{1}\frac{k_{+}k_{1-}}{kk_{1}}\right) \left(
1+s_{2}s^{\prime }\frac{k_{2+}k_{-}^{\prime }}{k_{2}k^{\prime }}\right)
\langle k_{1}s_{1}|K|k_{2}s_{2}\rangle .
\end{eqnarray}%
To evaluate the expression for $\langle VKV\rangle _{AV}$, we first use $%
\tilde{G}(k_{1}s_{1})\langle k_{1}s_{1}|J_{x}|k_{2}s_{2}\rangle \tilde{G}%
(k_{2}s_{2})$ for $\langle k_{1}s_{1}|K|k_{2}s_{2}\rangle $. Because $J_{x}$
is diagonal in $k$, ${\bf k}_{1}={\bf k}_{2}$, and the average of the
exponential factor leads to ${\bf k}={\bf k}^{\prime }$. Repeating this
procedure iteratively, we find that, like $\tilde{G}$, $K$ is diagonal in $k$%
. The matrix elements $\langle ks|VKV|ks^{\prime }\rangle $ may be evaluated
iteratively. We call $\langle ks|K|ks^{\prime }\rangle ^{\left( 0\right)
}=\langle ks|\tilde{G}J_{x}\tilde{G}|ks^{\prime }\rangle $ and note that $%
\langle ks|\tilde{G}J_{x}\tilde{G}|ks^{\prime }\rangle \propto k_{x}$ when $%
s=s^{\prime }$ and $\propto k_{y}$ when $s\neq s^{\prime }$ and keeping
terms which are even functions of $k_{x}$ and $k_{y}$ in the summation of
the equation. Then, by direct inspection, 
\begin{equation}
\langle ks|VKV|ks^{\prime }\rangle _{AV}=\frac{e\lambda ^{\prime }}{k}%
\left( k_{x}{\bf \sigma }_{z}-k_{y}{\bf \sigma }_{y} \right) \,,
\end{equation}%
where 
\begin{eqnarray}
e\lambda ^{\prime } &=&\frac{nV^{2}}{2L^{2}}\sum_{k}\frac{1}{k} \left(%
k_{x}\langle k+|\tilde{G}J_{x}\tilde{G}|k+\rangle -ik_{y}\langle k+|\tilde{G}%
J_{x}\tilde{G}|k-\rangle  \nonumber \right. \\
&& + \left. ik_{y}\langle k-|\tilde{G}J_{x}\tilde{G}|k+\rangle -k_{x}\langle k-|%
\tilde{G}J_{x}\tilde{G}|k-\rangle \right) .  \label{lamprime}
\end{eqnarray}%
Comparing Eq. (14) with Eq. (\ref{vertex}) and the definition of $J_{x}$,
Eq. (4), we find that $K\equiv \tilde{G}\tilde{J}_{x}\tilde{G}$ has the same
structure as $\tilde{G}J_{x}\tilde{G}$, where $\lambda $ in $J_{x}$ is
replaced by $\lambda +\lambda ^{\prime }$ in $\tilde{J}_{x}$. By replacing $%
\tilde{G}J_{x}\tilde{G}$ in Eq. (\ref{lamprime}) with $K=\tilde{G}\tilde{J}%
_{x}\tilde{G},$ we obtain a closed equation for $\lambda ^{\prime }$: 
\begin{equation}
\lambda ^{\prime }=\frac{nV^{2}}{4L^{2}}\sum_{k} \left\{ kbS_{1}+(\lambda
+\lambda ^{\prime })S_{0} \right\} ,
\end{equation}%
with $S_{1}\equiv \sum_{s}s\tilde{G}_{ks}\tilde{G}_{ks}$ \ and $S_{0}\equiv
\sum_{ss^{\prime }}\tilde{G}_{ks}\tilde{G}_{ks^{\prime }}.$ Here, we have
used relations $k_{x}^{2}=$ $k_{y}^{2}=k^{2}/2$ in the summation over $k.$

The conductivity now reads 
\begin{equation}
\sigma _{xx}=\frac{\hbar }{2\pi L^{2}}{\rm Tr}J_{x}\tilde{G}\tilde{J}_{x}%
\tilde{G},
\end{equation}%
where the tilde indicates substitution of $\lambda $ by $\lambda +\lambda
^{\prime },$ and $J_{x}\tilde{G}\tilde{J}_{x}\tilde{G}$ is a 2$\times 2$
matrix expressed in $s$-space. By carrying out trace, the conductivity
follows as $\sigma _{xx}=\sigma _{xx}^{+}+\sigma _{xx}^{-}$ with 
\begin{equation}
\sigma _{xx}^{\pm }=\frac{\hbar e^{2}}{4\pi L^{2}}\sum_{k} \left\{ \left(
b\pm \frac{\lambda }{k}\right) \left( b\pm \frac{\lambda +\lambda ^{\prime }%
}{k} \right) k^{2}\tilde{G}_{k\pm }\tilde{G}_{k\pm }+\lambda (\lambda
+\lambda ^{\prime })\tilde{G}_{k\mp }\tilde{G}_{k\pm } \right\} .
\label{sigma}
\end{equation}%
By using the unitary matrix $U,$ the matrix representation of the
conductivity in $s$-space can be transformed into that in the original Pauli
spin-space as ${\hat{\sigma}}_{xx}=\hbar UJ_{x}\tilde{G}\tilde{J}_{x}\tilde{G%
}U^{\dagger }/2\pi L^{2},$ with $\hat{\sigma}_{xx}^{\uparrow \uparrow },\hat{%
\sigma}_{xx}^{\uparrow \downarrow }$ {\it etc}. By taking the spin trace{\rm %
\ }of{\rm \ }${\hat{\sigma}}_{xx},$ the relation $\sigma _{xx}^{+}+\sigma
_{xx}^{-}=\hat{\sigma}_{xx}^{\uparrow \uparrow }+\hat{\sigma}%
_{xx}^{\downarrow \downarrow }$ follows naturally. We observe that $\hat{%
\sigma}_{xx}^{\uparrow \uparrow }=\hat{\sigma}_{xx}^{\downarrow \downarrow }$
and that the non-diagonal elements of the conductance tensor in the original
spin-space $\hat{\sigma}_{xx}^{\uparrow \downarrow }$ vanish identically by
parity. These results prove that the current excited by the electric field
is not spin polarized. The expression for $\sigma _{yy}$ can be derived
analogously. We also find that the non-diagonal (Hall) conductivity $\sigma
_{xy}$ vanishes by symmetry.

In calculating the vertex correction and conductivity we encounter integrals
over the momentum, which may easily evaluated by an approximation in which
the life time broadening of the density of states is neglected, 
\begin{equation}
\tilde{G}_{ks}^{R}\tilde{G}_{ks}^{A}=\frac{2\pi \tau }{\hbar }\delta
(\epsilon -E_{ks}),  \label{ints}
\end{equation}%
where the life time $\tau $\ is defined $\Sigma =-i{\rm sign}(\eta )\hbar
/2\tau $ or ${\tau }={\hbar /2}\pi nV^{2}D$ with the 2DEG density of states
per spin $D=m/2\pi \hbar ^{2}$. We then obtain $\lambda ^{\prime }=-\lambda
, $ and 
\begin{equation}
\sigma _{xx}=\frac{2e^{2}n_{0}\tau }{m}+2e^{2}\tau D\lambda ^{2},
\end{equation}%
where $n_{0}$ is the number of electrons per spin. The conventional Drude
conductivity (first term) is increased by the spin-orbit interaction. Since
the sign of the coupling constant $\lambda $ is irrelevant, the enhancement
term must be of even order in $\lambda $ . The result $\lambda ^{\prime
}=-\lambda $ shows that the vertex function $\tilde{J}_{x}$ is diagonal in
spin space\cite{schwab}. \ In the case of $\lambda =0$, the vertex
correction in the ladder approximation vanishes identically due to the
isotropic scattering.

It is important to distinguish the spin-polarized currents computed above
from the spin-accumulation which is excited by the applied field $E$, which
in linear response is given by: 
\begin{equation}
\left\langle {\bf s}\right\rangle =\hbar {\rm Tr}\text{\boldmath$\sigma $}%
\left\langle G^{A}J_{x}G^{R}+G^{R}J_{x}G^{A}\right\rangle _{AV}E.
\end{equation}%
In $s$-space, each component is given as 
\begin{equation}
\left\langle s_{i}\right\rangle =\hbar {\rm Tr}U^{\dagger }\sigma _{i}U%
\tilde{G}\tilde{J}_{x}\tilde{G}E.
\end{equation}%
We find $\langle s_{z}\rangle =\langle s_{x}\rangle =0$, but 
\begin{equation}
\left\langle s_{y}\right\rangle =-\frac{e\hbar }{2L^{2}}\sum_{k} \left\{%
bkS_{1} + (\lambda +\lambda ^{\prime })S_{0} \right\} E.
\end{equation}%
The expression is simplified as 
\begin{equation}
\left\langle s_{y}\right\rangle =e4\pi \tau D\lambda E,
\end{equation}%
by using the approximation (\ref{ints}). The spin accumulation is aligned to
the pseudo-magnetic field of the spin-orbit interaction and its magnitude is
proportional to the applied electric field within the linear response regime.%
{\it \ }The magnitude of the spin accumulation may be estimated as%
\begin{equation}
\left| \left\langle s_{y}\right\rangle \right| /D=2.5\left[ \frac{\lambda }{%
10^{-11}{\rm eVm}}\right] \left[ \frac{eE}{10\cdot {\rm KeV/m}}\right] \left[
\frac{\mu }{10^{5}{\rm cm}^{2}{\rm V}^{-1}{\rm s}^{-1}}\right] {\rm meV}
\end{equation}%
With{\it \ }$\lambda =10^{-11}$ eVm \cite{nitta} the splitting of the
chemical potential thus amounts to a significant 2.5 meV for typical
experimental parameters of the applied field and the mobility $\mu $.

We have thus formulated the conductivity tensor and spin accumulation on an
equal footing using linear response theory. The conductivity is found to be
enhanced by the spin-orbit interaction and isotropic in spin space, i.e., $%
\hat{\sigma}_{xx}^{\uparrow \uparrow }=\hat{\sigma}_{xx}^{\downarrow
\downarrow }$, and $\hat{\sigma}_{xx}^{\uparrow \downarrow }=\hat{\sigma}%
_{xx}^{\downarrow \uparrow }=0$. We have proven that an electric field
induced a spin-polarized density \cite{edelstein}, but not a spin-polarized
current. The recent discovery of the spin-galvanic effect, i.e. that a
magnetization along $y$\ direction induces an electric current \cite%
{ganichev} is reciprocal to the current induced spin accumulation. We may
conclude that his current can also not be spin-polarized.

The result that the conductivity is spin-isotropic implies that the spin
accumulation in ferromagnet(F)/2DEG hybrids cannot be detected in
two-terminal configuration with one ferromagnetic contact. A single source
or drain ferromagnetic contact does not modify the global transport
properties in the diffusive regime, because the contacts, which connect the
reservoir distribution functions to the semiconductor ones, are not affected
by a magnetization reversal\cite{molenkamp}. A phenomenological theory\cite%
{johnson,silsbee} is at odds with this conclusion. Microscopically, we trace
the matrix character of the current operator as the culprit of this
disagreement\cite{molenkamp}. Experiments on F/2DEG systems\cite{hammar},
which were supported by that theory\cite{johnson,silsbee}, were challenged
by Monzon $et~al.$ \cite{monzon} and van Wees\cite{wees}, who suspected that
the measured effects were due to local Hall voltages caused by fringe fields
near the ferromagnetic contacts. These \cite{hammar} and subsequent
experiments\cite{hammar2,hammar3} should perhaps be reconsidered in the
light of the present theoretical results.

The present results are related but different from the spin-Hall effect
discussed by Zhang\cite{zhang} who dealt with a ferromagnetic metal
thin-film. He suggested to measure the spin-accumulation excited by the
electric current in a three-terminal configuration, however, which is also
an option for the 2DEG, since we find the spin accumulation signal to be
quite significant.

Spin-polarized transport can be detected in a F/2DEG/F configuration with
two ferromagnetic contacts as studied by Pareek and Bruno \cite{pareek}{\it %
. W}e point out that the spin dependent conductances $\Gamma _{\uparrow
\uparrow }$ and $\Gamma _{\uparrow \downarrow }$ in Ref. \cite{pareek} must
not be confused with the spin dependent conductivities $\hat{\sigma}%
_{xx}^{\uparrow \uparrow }$ and $\hat{\sigma}_{xx}^{\uparrow \downarrow }$
defined above. The former conductances are defined as $\Gamma _{\uparrow
\uparrow (\downarrow )}=(e^{2}/h){\rm Tr}{\bf t}_{\uparrow \uparrow
(\downarrow )}{\bf t}_{\uparrow (\downarrow )\uparrow }^{\dagger }$, where $%
{\bf t}$ is a transmittance matrix. Whereas $\hat{\sigma}_{xx}^{\uparrow
\downarrow }=0$ as shown above, in general $\Gamma _{\uparrow \downarrow
}\neq 0$. It is possible to decompose the conductivity $\hat{\sigma}%
_{xx}^{\uparrow \uparrow }$ by ``cutting lines''\ in the conductivity
diagram such that $\hat{\sigma}_{xx}^{\uparrow \uparrow }=\hat{\sigma}%
_{xx}^{\uparrow \uparrow \uparrow \uparrow }+\hat{\sigma}_{xx}^{\uparrow
\downarrow \downarrow \uparrow },$ which is simplified by the spin-diagonal
vertex function in the ladder approximation. The two components are then
given as 
\begin{equation}
\hat{\sigma}_{xx}^{\uparrow \uparrow \uparrow \uparrow }=\frac{\hbar e^{2}}{%
16\pi L^{2}}\sum_{k,s}(b^{2}k^{2}+sb\lambda k)\tilde{G}_{s}(\tilde{G}_{+}+%
\tilde{G}_{-})
\end{equation}%
and 
\begin{equation}
\hat{\sigma}_{xx}^{\uparrow \downarrow \downarrow \uparrow }=\frac{\hbar
e^{2}}{16\pi L^{2}}\sum_{k,s}s(b^{2}k^{2}+sb\lambda k)\tilde{G}_{s}(\tilde{G}%
_{+}-\tilde{G}_{-}).
\end{equation}%
We did not find a simple relation as Eq. (20) but in general $\hat{\sigma}%
_{xx}^{\uparrow \downarrow \downarrow \uparrow }\neq \hat{\sigma}%
_{xx}^{\uparrow \uparrow \uparrow \uparrow }$, because in the limit $\lambda
=0$, $\hat{\sigma}_{xx}^{\uparrow \downarrow \downarrow \uparrow }=0,$
whereas $\hat{\sigma}_{xx}^{\uparrow \uparrow \uparrow \uparrow }$ tends to
the Drude conductivity. This conclusion appears to be at odds with the
numerical findings of Pareek and Bruno that in the limit of long samples $%
\Gamma _{\uparrow \uparrow }\sim \Gamma _{\uparrow \downarrow }$.

In summary, we derived explicit expressions for the conductivity tensors and
spin accumulation of a Rashba 2DEG with isotropic scattering centers, taking
into account the vertex correction in the ladder approximation. The
diffusive conductivity limited by non-magnetic impurity scattering is not
spin dependent, although the applied bias does excite a spin accumulation.

We acknowledge helpful explanations by Yuli Nazarov of his early work\cite%
{levitov}. The present research is supported by NEDO International Joint
Research Grant, ``Nano-scale Magnetoelectronics'' and the FOM. JI thanks the
hospitality at Delft University of Technology where this work started and
acknowledges financial support from the CREST projects of Japan ``Photonic
function of materials with nanometer-sized structures'' and ``Spin-injection
as a base of the spin-charge coupled electron devices'', and useful
discussions with B. van Wees, P. Bruno and H. Itoh. LWM acknowledges
sponsoring by the DARPA SPINS program, DFG(SFB(410)) and the German BMBF,
and GEWB support by the FOM as well as the hospitality of dr. Y. Hirayama at
the NTT Basic Research Laboratories.

\end{document}